\documentclass[12pt]{article}
\usepackage{graphicx}
\usepackage{amssymb}
\input epsf

\def\beq{\begin{equation}}
\def\eeq{\end{equation}}
\def\phid{\phi^{\dagger}}
\def\pd{\partial}

\def\phid{\phi^{\dagger}}
\title{\bf Finite nuclei in the reggeon "toy model"}
\author{M.A.Braun\\
{\it Sankt-Petersburg State University, Russia}}

\begin{document}
\maketitle
\begin{abstract}
Hadron-nucleus amplitudes at high energies are studied in the "toy" Regge model in zero transverse dimension
for finite nuclei, when the standard series of fan diagrams is converted into a finite sum and looses physical
sense at quite low energies. Taking into account all the loop contributions by numerical methods we find a physically
meaningful amplitudes at all energies. They practically coincide with the amplitudes for infinite nuclei.
A surprizing result is that for finite nuclei and small enough triple pomeron coupling the infinite series of fan diagrams
describes the amplitude quite well in spite of the fact that in reality the series should be cut and as such deprived of
any physical sense at high energies.
\end{abstract}

\section{Motivation}
At high energies in the framework of the perturbative Quantum Chromodynamics strong interactions are mediated by
the exchenge of hard pomerons, which are non-local entities propagating according to the BFKL equation and splitting
into two or merging from two to one with the known triple pomeron vertex. Neglecting pomeron loops
and choosing the projectile to
have a short range and the target to be a heavy nucleus one comes to the well-known Balitski-Kovchegov (BK) equation,
which sums
pomeron fan diagrams going from the projectile towards the target ~\cite{bal,kov}. This equation with certain degrees of
sophistication, including
higher orders and running coupling, is widely used in applications with very positive results. However from the start
it is clear that summing all fans this equation neglects the obvious limitation that the nucleus in fact is finite,
so that the number of splittings in fans is restricted. It is not clear how this restriction affects the resulting
amplitude. This fact cannot be too small. For instance,
taking instead of a heavy nucleus a light one, say the deuteron, one sees that the results become drastically different,
since then the amplitude essentially reduces to the exchange of two pomerons ang grows correspondingly at high energies.

This problem is difficult to study in the framework of the fully-fledged QCD pomeron theory. However very long ago
a description
of the hA interaction by means of the sum of fan diagrams was proposed by A.Schwimmer in  the reggeon field
theory with a local supercritical pomeron ~\cite{schwimmer}. There one easily obtains a solution of the equation,
which sums all fans and
is basically similar to the BK equation. Taking into account that the momenta transferred to the nucleus
are small one gets for the hA amplitude at given rapidity $y$ and fixed impact parameter $b$
\beq
{\cal A}(b)=\frac{g^2AT(b)e^{\mu y}}{1+\frac{AgT(b)\lambda}{\mu}\Big(e^{\mu y}-1\Big)}.
\label{schw}
\eeq
Here $\mu$ is the pomeron intercept minus unity, assumed to be positive, $T(b)$ is the nuclear profile function
normalized to unity, $\lambda>0$ is the triple pomeron coupling with the opposite sign and
$g$ is the pomeron-nucleon coupling. This old formula possesses some nice features. In particular the Schwimmer amplitude
goes to a constant value at very high energies, implying that at such energies the nucleus behaves like a (grey) disk,
which more or less agrees with the later more sophisticated treatments. We shall be interested in not so much of its
physical applications but rather as a tool to study our problem: how the limitations on the number of splitting coming from
the actual finiteness of $A$ change the result.

Having the explicit solution (\ref{schw}) this is quite trivial to see. Let
\beq
z=\frac{AgT(b)\lambda}{\mu}\Big(e^{\mu y}-1\Big),
\eeq
so that the Schwimmer amplitude is just
\beq
{\cal A}(b)=\frac{gAT(b)e^{\mu y}}{1+z}.
\label{schw1}
\eeq
For a finite nucleus with atomic number $A$ obviously we get just $A$ first terms of the
expansion of (\ref{schw1}) in powers of $z$
\beq
{\cal A}_A(b)=gAT(b)e^{\mu y}\sum_{n=0}^A(-z)^n=gAT(b)e^{\mu y}\frac{1-(-z)^{A+1}}{1+z}.
\label{schwf}
\eeq
Here and in the following we denote with subindex $A$ the amplitudes which refer to finite nuclei of atomic number $A$,
leaving the amplitude for infinite nucleus without any subindex. One immediately sees that if $z<1$ then the series in
(\ref{schwf}) is convergent and the error in using the infinite nuclei instead of the finite (and phyiscal) is exponentially
small. On the other hand, if $z>1$ the series (\ref{schwf}) is divergent and the Schwimmer formula for infinite nucleus
has nothing in common with the real amplitude for finite nucleus.

Condition of validity of the fan diagram amplitude $z<1$ translates into the restriction on the highest rapidity where
this amplitude has the physical meaning
\beq
e^{\mu y}<1+\frac{\mu}{AgT(b)\lambda}.
\label{cond}
\eeq
If the internucleon distance in the nucleus is $R_0$ then crudely estimating we have
\[g\sim R_0,\ \ AT(b)=A^{1/3}/R_0^2\]
and (\ref{cond}) gives
\beq
y<\frac{1}{\mu}\ln\Big(1+A^{-1/3}\frac{\mu R_0}{\lambda}\Big).
\label{cond1}
\eeq
With fixed $\lambda$ independent of $A$ and
large $A>>1$ this degenerates into
\beq
y<A^{-1/3}\frac{\mu R_0}{\lambda}<<1,
\eeq
which has little physical meaning. The only possibility to have some sense
for the Schwmimmer amplitude for realistic nuclei is to choose $\lambda$
extremely small to compensate factor $A^{-1/3}$ in (\ref{cond1}).
Otherwise this formula is just an analytic continuation of the
physical amplitude having no relation to  reality.

This problem is  of course not a new one and is not restricted to only fan
diagrams. In a simpler case  of the scattering of a hadron on a nuclear target in the Glauber approximation
one gets the amplitude
\beq
i{\cal A}_A(b)=\Big(1+ia(y)T(b)\Big)^A-1,
\label{glau}
\eeq
where $a$ is the forward proton-proton scattering amplitude.
In the limit $A\to\infty$ and $T(b)\propto A^{-2/3}$ one gets the standard expression
\beq
{\cal A}(b)=1-e^{iAT(b)a(y)},
\eeq
which is quite atttractive, since it is explicitly unitary. However  expression
(\ref{glau}) is unitary only while $|a(y)T(b)|<1$. This is always so when the proton-proton
scattering amplituide is unitary itself. However if one takes for $a(y)$ the amplitude corresponding
to the exchange of a supercritical pomeron and so rising with $y$ as $\exp (\mu y)$ then the expression
for finite nuclei looses sense for high enough $y$. The limiting expression for infinite nuclei
preserves its unitary character but its relation to the physical amplitude becomes lost.

Observing these examples we may conclude that in both cases, fans and Glauber, the origin of the difficulty
lies in the wrong behaviour of  the elementary proton-proton scattering amplitude (pomeron propagator)
growing exponentially with rapidity
and violating unitarity. This wrong behaviour is possibly cured by inclusion of contributions from pomeron loops.
We cannot be sure that this happens in the reggeon field theory, which lies at the basis of the Schwimmer formula
(\ref{schw}). Still less is known about the behaviour of the propagator of the non-local
pomeron in the perturbative QCD. However there is a simple model where calculation of all loop contributions
is possible. This is a reggeon "toy model" in the zero-dimensional transverse space. Having rapidity as the only
variable it actually reduces to a sort of quantum mechanics with a non-Hermithean interaction. This theory has
been extensively studied in the past ~\cite{amati1}-~\cite{ciafaloni2} and in the limit $\lambda\to 0$
it was shown analytically that inclusion of
loops makes the pomeron propagator vanish in the high-energy limit. More recently it was considered in
\cite{bondarenko,bravac} and in ~\cite{bravac} a calculational
technique was elaborated which allowed to numerically sum all contributioins for arbitrary values of parameters.
In this paper we apply this technique to study the behaviour of the hA amplitude in the model for finite nuclei
and establish the relation between the sum of fan diagrams and the full amplitude in this case.

Note that the fan amplitude in the toy model is identical to the more physical Schwimmer amplitude (\ref{schw}).
The simplification of neglecting the transverse space is felt only in the loop diagrams, which in the toy
model are certainly different from the ones in the reggeon field theory with transverse dimensions.
Still we expect that lessons known in the study of the toy model will be instructive to consider the situation
in more physical theories including the perturbative QCD.

\section{The toy model}
in this section we briefly recapitulate the
definition and propreties of the toy model indispensable for
our study, referring the reader either to old papers  ~\cite{amati1}-~\cite{ciafaloni2}
 or to the comparatively recent paper  ~\cite{bravac}.
The toy model is the Regge-Gribov theory of a  pomeron field $\phi(y)$ depending only on rapidity $y$
(zero-dimensional transverse space), which may be defined by the functional integral
\beq
Z=\int D\phi D\phid e^{-S},\ \ S=\int dy {\cal L},
\label{funint}
\eeq
where
\beq
{\cal L}=\frac{1}{2}(\phid\phi_y-\phid_y\phi)-\mu\phid\phi
+i\lambda\phid\phi(\phid+\phi).
\label{lagrangian}
\eeq
Here $\mu$ is the pomeron intercept ($\alpha(0)-1$). For the
supercritical pomeron $\mu>0$. Triple pomeron coupling constant
$\lambda$ is also positive. The functional integral (\ref{funint}) converges for $\mu<0$ (subcritical pomeron). But in
the physically interestng
case when $\mu>0$ the integral does not exist. Then in fact it only serves to introduce perturbative diagrams in the
Regge-Gribov approach.

One can pass to an alternative, Hamiltonian formalism, which  reproduces the perturbative diagrams but is free
from the restriction $\mu<0$. It is based on a quasi-Schroedinger equation in rapidity for the wave function $\Psi(y)$
\beq
\frac{d\Psi(y)}{dy}=-H\Psi(y)
\eeq
with the Hamiltonian $H$ which can be chosen to be real
\beq
H=\mu u v-\lambda u(u+v)v
\label{Huv}
\eeq
and is a function of two
operators $u$ and $v$, which are anti-Hermithean to each other
\beq
u^\dagger=-v,\ \ v^\dagger=-u
\eeq
and satisfy the commutation relation
\beq
[v,u]=-1.
\label{commrel}
\eeq
Operators $u$ and $v$ have the meaning of creation and annihilation operators of the pomeron respectively.
The vacuum state
$\Psi_0$, normalized to unity, satisfies $v\Psi_0=0$.
All other states are built from $\Psi_0$ by application of some number
of operators $u$. The transition amplitude from the initial state
$\Psi_i$  at rapidity $y=0$ to the final state $\Psi_f$
at rapidity $y$ is given by
\beq
iA_{fi}=\langle \Psi_f|\Psi_i(y)\rangle,\ \
\Psi_i(y)=e^{-Hy}\Psi_i.
\label{mel}
\eeq
The amplitude $A_{fi}$ is imaginary positive so that the matrix element on
the right-hand side of (\ref{mel}) is negative.
Some care should be taken to express the initial and final
scattering states $\Psi_i(u)$ and $\Psi_f(u)$  via creation operators. We take them also to be real. Assuming that the
initial state representing a heavy nucleus with $A\to \infty$ has an eikonal structure we take
\beq
\Psi_i(u)=\Big(1-e^{-g_iu}\Big)\Psi_0,
\label{ini}
\eeq
where $g_i$ is a positive coupling constant with the initial nucleus. It is important that the final state
should be taken not as an immediate copy of (\ref{ini}) (with a maybe different coupling constant) but with
an additional change $u\to -u$
\beq
\Psi_f(u)=\Big(1-e^{+g_fu}\Big)\Psi_0.
\label{fin}
\eeq
As we shall see this immediately follows from the form of the amplitude at $y=0$.
So for the scattering of two nuclei we get the amplitude in terms of purely real quantities
\beq
i{\cal} A=
\langle
\Big(1-e^{g_fu}\Big)\Psi_0|e^{-Hy}|\Big(1-e^{-g_iu}\Big)\Psi_0\rangle=
\langle\Big(1-e^{-g_fv}\Big)e^{-Hy}\Big(1-e^{-g_iu}\Big)\rangle.
\eeq
In the last formula the vacuum matrix element is implied.
Since $H\Psi_0=0$ the term independent of $g_i$ and $g_f$ vanishes, so
that
we can also write
\beq
iA_{fi}=
-\langle e^{-g_fv}e^{-Hy}\Big(1-e^{-g_iu}\Big)\rangle=
-\langle\Psi_0|e^{-g_fv}F_i(y,u)\Psi_0\rangle,
\label{ampli1}
\eeq
where $F_i(y,u)$ is the operator which creates the evolved initial state.
It satisfies the equation
\beq
\frac{\pd F_i(y,u)}{\pd y}=-H(u,v)F_i(y,u)
\label{evol}
\eeq
with the initial condition
\beq
F_i(0,u)=1-e^{-g_iu}.
\label{initf}
\eeq
The commutation relation (\ref{commrel}) allows to represent
\beq
v=-\frac{\pd}{\pd u}
\label{vrepr}
\eeq
and then (\ref{ampli1}) implies that to find the amplitude one has to substitute
$u$ by $g_f$ in $F_i(y,u)$
\beq
iA_{fi}=-F_i(y,g_f).
\eeq

At $y=0$ this gives the initial amplitude
\beq
{\cal A}=i\Big(1-e^{-g_fg_i}\Big)
\eeq
in clear correspondence with the nucleus-nucleus amplitude in the so-called optical approximation.
Should we take the final state without reversing the sign of $u$ we would get the sign plus
in the exponent in obvious contradiction with the optical amplitude.

Taking the complex conjugate  of (\ref{ampli1}) we find
\beq
-iA^*_{fi}=
\langle \Big(1-e^{g_iv}\Big)e^{-H^{\dagger}y}\Big(1-e^{g_fu}\Big)\rangle
=iA_{if}(\lambda\to -\lambda,g_{i(f)}\to -g_{f(i)}).
\eeq
Having in mind that the amplitude is pure imaginary,
we see that interchanging the target and projectile leads to the overall change of sign
$u\to -u$. However this will not change the amplitude. Indeed after evolution  we
shall get function $F_i(y,-u)$. But the change $u\to -u$ in the final state requires that now we have
to substitute $u$ by $-g_f$ so that the result will be the same $F_i(y,g_f)$ as for the direct transition.
So  the interchange of the target and projectile does not change the amplitude.

\section{Numerical studies}
Calculation of the scattering amplitude reduces to the solution of the differential equation
in two variables, $y$ and $u$
\beq
\frac{\pd F_i(y,u)}{\pd y}=\Big(\mu u \frac{\pd}{\pd u}-\lambda u^2\frac{\pd}{\pd u}+
\lambda u\frac{\pd^2}{\pd u^2}\Big)F_i(y,u),
\label{evol1}
\eeq
which determines evolution in $y$ of function $F(y,u)$ initally given at $y=0$:
$F(y,u)|_{y=0}=F_0(u)$.
Note that apart from the chosen $A$ the amplitude depends on $y$ and two parameters $\mu$ and $\lambda$. From the form
of the Eq. (\ref{evol1}) it folows that this last three variables are combined in two: the scaled rapidity $\bar{y}=\mu y$
and ratio $\rho=\mu/\lambda$. So one can explore the whole domain of rapidities $y$ and values of $\mu$ and $\lambda$
by limiting $\bar{y}\leq \bar{y}_{max}$
and changing values of $\rho$ appropriately.

Equation (\ref{evol1}) can be solved anaytically only in the case when one drops the term with the second derivative,
which describes fusing of pomerons. The remaining equation with only the first derivatives describes propagating pomerons
and their
consecutive splittings, that is fan diagrams. Its solution can be easily obtained to be ~\cite{bravac}
\[ F_0(y+z),\ \ z=\frac{1}{\mu}\ln\frac{u}{u-\rho},\ \ \rho=\frac{\mu}{\lambda}.\]
For the fan amplitude one chooses the initial state to be a single pomeron $F_0(u)=g_iu$ to find
\beq
F^{fan}(y,u)=\frac{g_iue^{\mu y}}{1+\frac{u}{\rho}\Big(e^{\mu y}-1\Big)}.
\label{fan}
\eeq
The amplitude itself is obtained from (\ref{fan}) by putting
$u=g_f$ and multiplcation by $i$ (it is essentially identical with the Schwimmer amplitude (\ref{schw}))

With the second derivative term included, solution of Eq. (\ref{evol1}) gives the complete amplitude
with all tree diagrams and loop diagrams taken into account. Note that for nucleus-nucleus scattering the
set of tree diagrams is much wider than the the set of fan diagrams. Unfortunately Eq. (\ref{evol1}) in this case
cannot be solved analytically (Note, however some important estimates at very small $\lambda$ in older papers
 ~\cite{amati1}-~\cite{ciafaloni2}).
So one is compelled to recur to numerical methods for the solution of Eq. (\ref{evol1}). In ~\cite {bravac} it was found that the
most straightforward approach of evolving the initial fuction in rapidity by the Runge=Kutta method proved to be quite
feasible, provided the step in $y$ is small enough and correlated wth the step in $u$. In our present calculation
we find good convergence  with $\Delta y=5.10^{-7}$, $\Delta u=1.10^{-2}$ and the interval in $u$ taken as
$0<u<20$. Further diminishing of $\Delta y$ or $\Delta u$ or raising the maxinum value of $u$ have been found to produce
no change whatsoever.

Calculations in  ~\cite{bravac} pursued a somewhat restricted goal to only illustrate the feasibility of the
numerical approach and see the limiting behaviour of the propagator and hA amplitude in the limit of very high rapidities.
Here we study the $A$ dependence of the hA amplitude having in mind finite nuclei of different atomic numbers and
comparison with the results for infinite nuclei. In the standard eikonal picture the effective coupling to the nucleus
grows as $A^{1/3}$, which comes from the product $AT(b)$. Accordingly we take for the nucleus $g_f=A^{1/3}$ and for the
nucleon $g_i=1$ for simplicity. The hA amplitude for the infinite nucleus is then found as explained earlier.
At $y=0$ we start from $F_0(u)=u$, evolve this function according to Eq. (\ref{evol1}) and take the final function
$F(y,u)$ at $u=g_f$.

However the main purpose of our calculations is to find what will happen when we consider realisitic nuclei with finite
atomic numbers $A$. To pass to finite nuclei we  change the eikonal amplitude for the infinite nucleus
to its standard Glauber form for the finite nucleus with atomic number $A$:
\beq
1-e^{-g_fu}\to 1-\Big(1-\frac{g_f}{A}u\Big)^A.
\label{inigla}
\eeq
If we change correspondingly the final state in the matrix element for the amplitude then after we evolve
the initial function $F_0(u)=u$ to the desired rapidity the amplitude will be given by
\beq
{\cal A}_A=\Big(1+\frac{g_f}{A}\frac{\pd}{\pd u}\Big)^AF(y,u)\Big|_{u=0}.
\eeq
For not very small values of $A$ numerical calculations of this expression are hardly feasible due to necessity to find
high-order derivatives.

So instead we use the discussed symmetry under the interchange of the projectile and target and calculate the inverse
amplitude with the initial state represented by the finite nucleus and the final one by the proton, that is
\beq
F_0(u)=1-\Big(1-\frac{g_f}{A}u\Big)^A,\ \ F_f(u)=-u.
\eeq

Our calculations were performed for the interval $0\leq\bar{y}\leq 5$ of the scaled rapidity for three values of
$\rho=10,2$ and $0.5$.
We recall that greater values of $\rho$ correspond to smaller values of the triple pomeron coupling $\lambda$.
Taking $\mu=0.1$ in more or less corresondence with the soft pomeron properties our amplitudes are found at rapidities
up to 50. The values of $\lambda$ studied are then $\lambda=0.01$ for $\rho=10$, $\lambda=0.05$ for $\rho=2$
and $\lambda=0.2$ for $\rho=0.5$.

\begin{figure}[ht]
\includegraphics[width=6 cm]{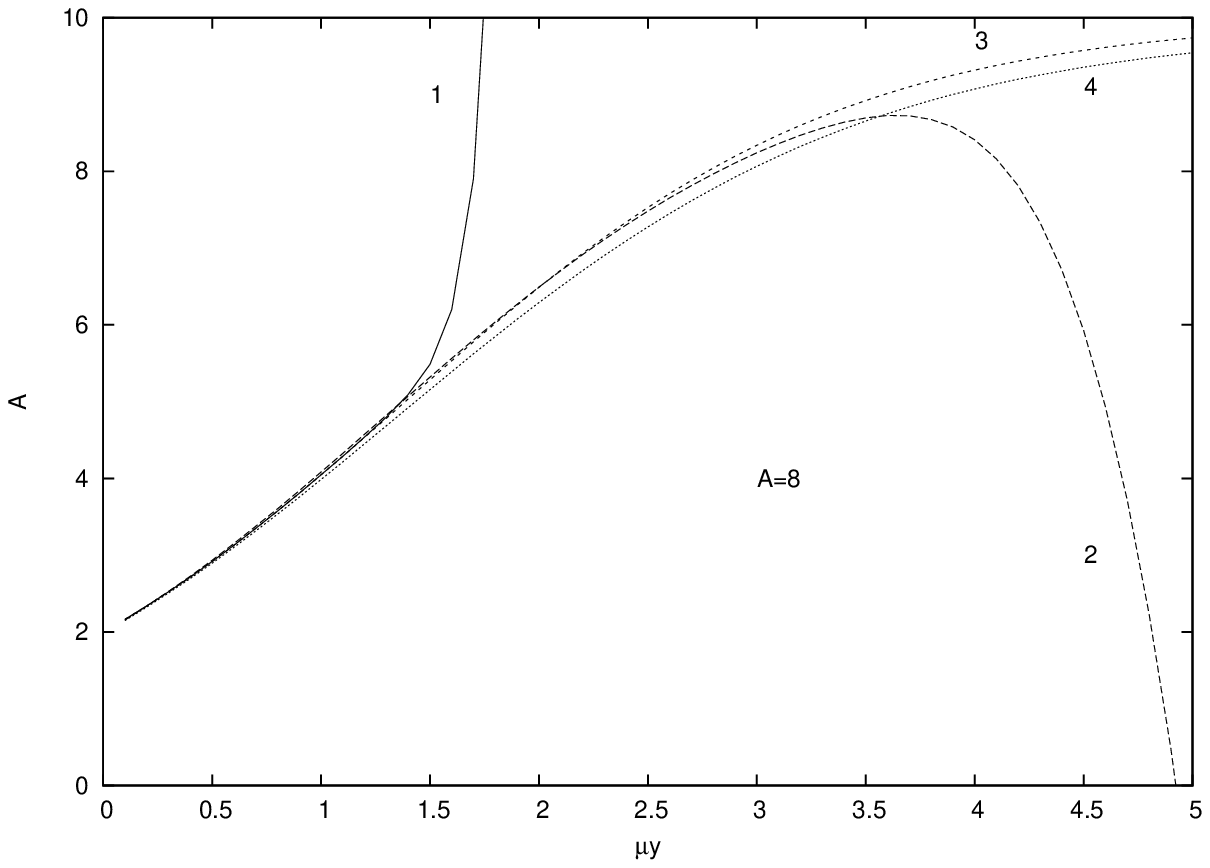}
\includegraphics[width=6 cm]{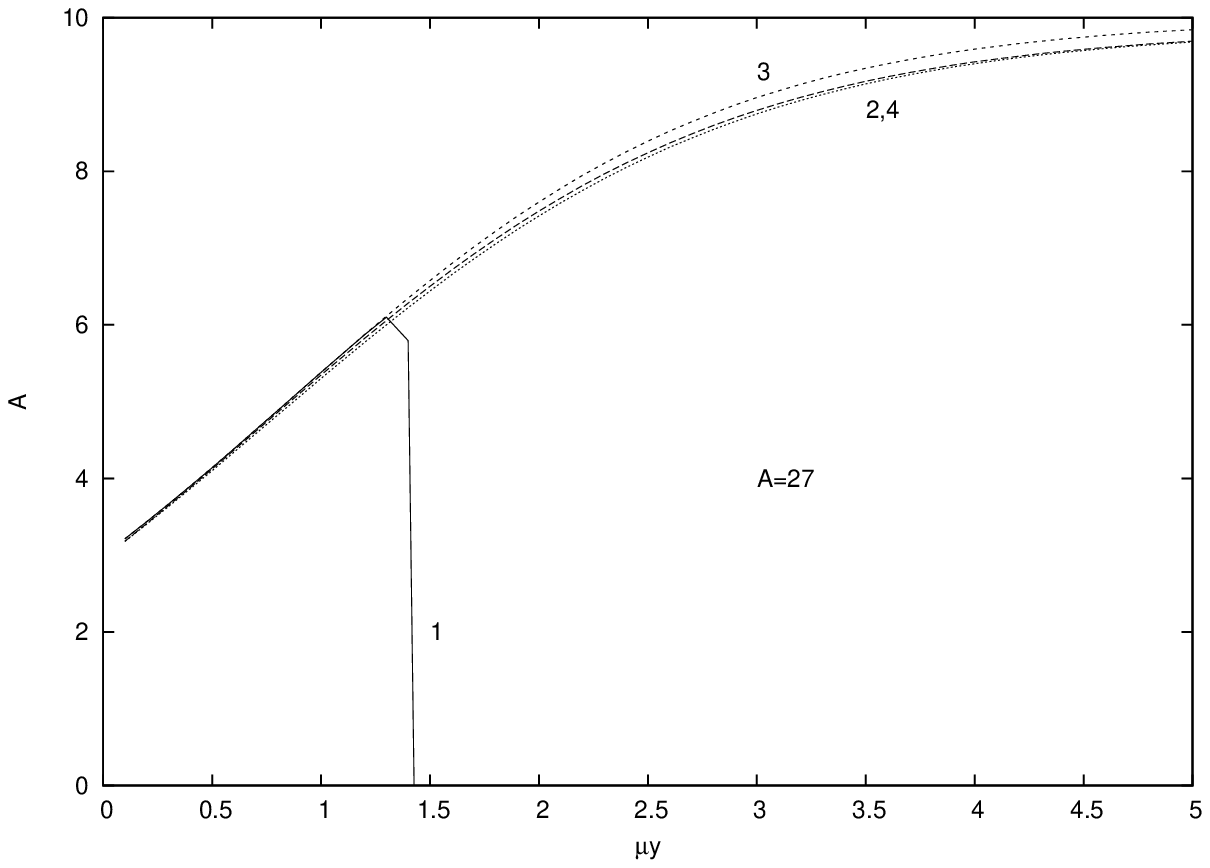}
\\
\vspace* {1 cm}
\includegraphics[width=6 cm]{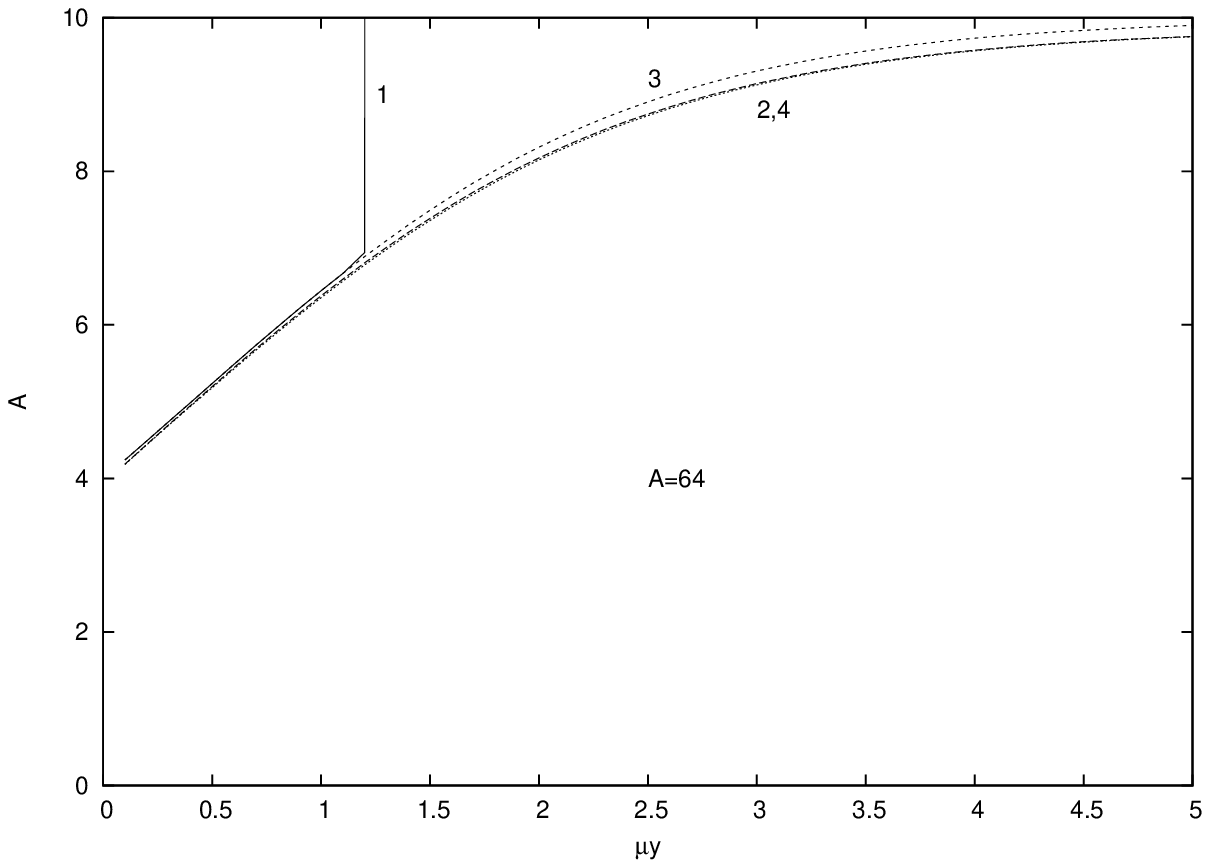}
\includegraphics[width=6 cm]{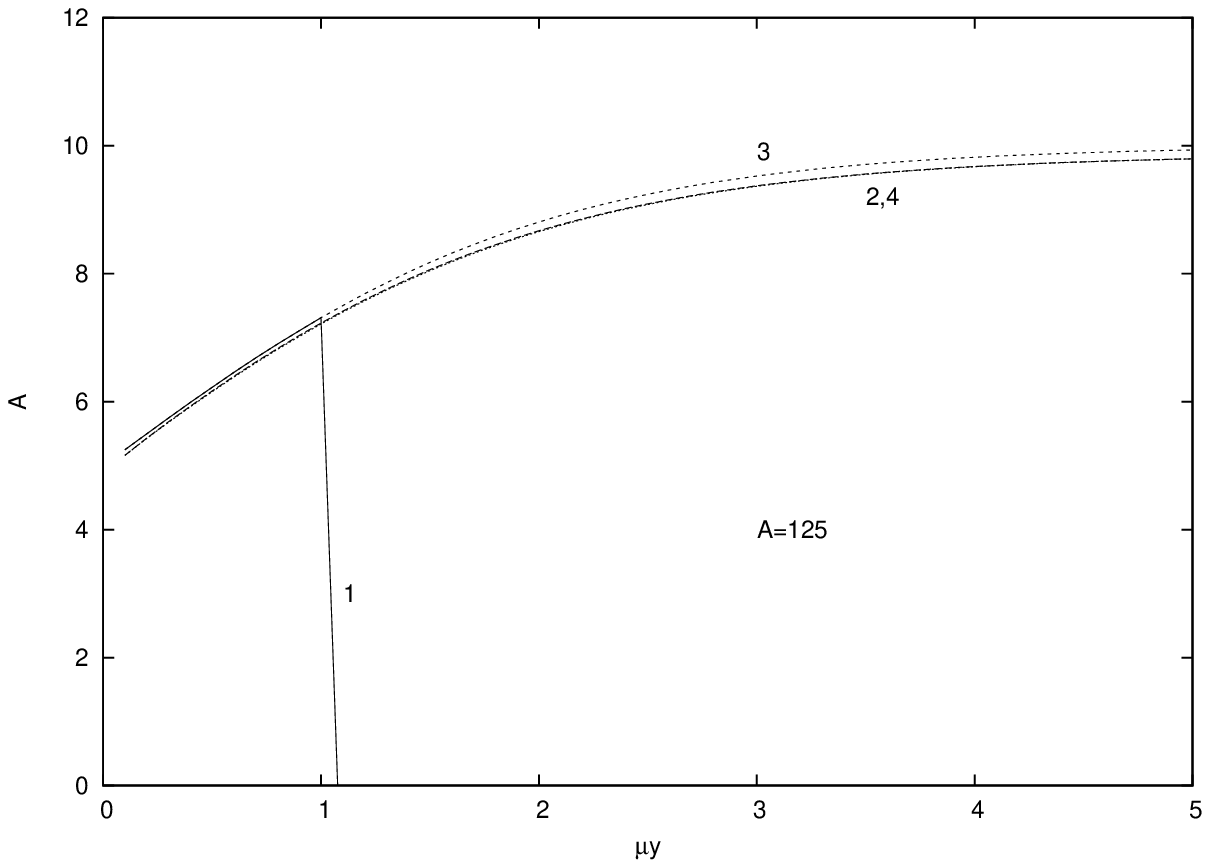}
\caption{hA amplitudes for $\rho=10$. Curves show:  fan amplitudes for finite nuclei (1),
full amplitude with loops for finite nuclei (2), infinite fan amplitudes (3),
full amplitudes for infinite nuclei (4)}
\label{fig1}
\end{figure}

\begin{figure}[ht]
\includegraphics[width=6 cm]{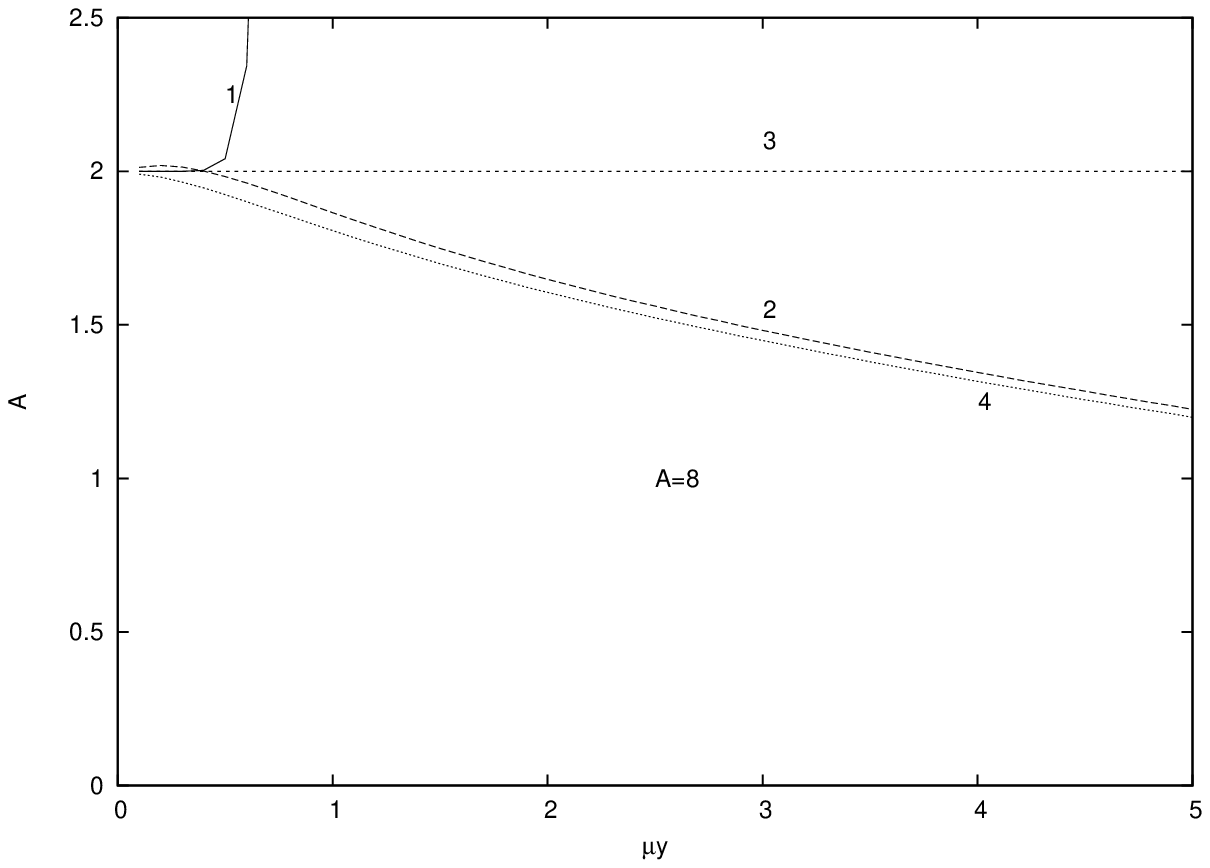}
\includegraphics[width=6 cm]{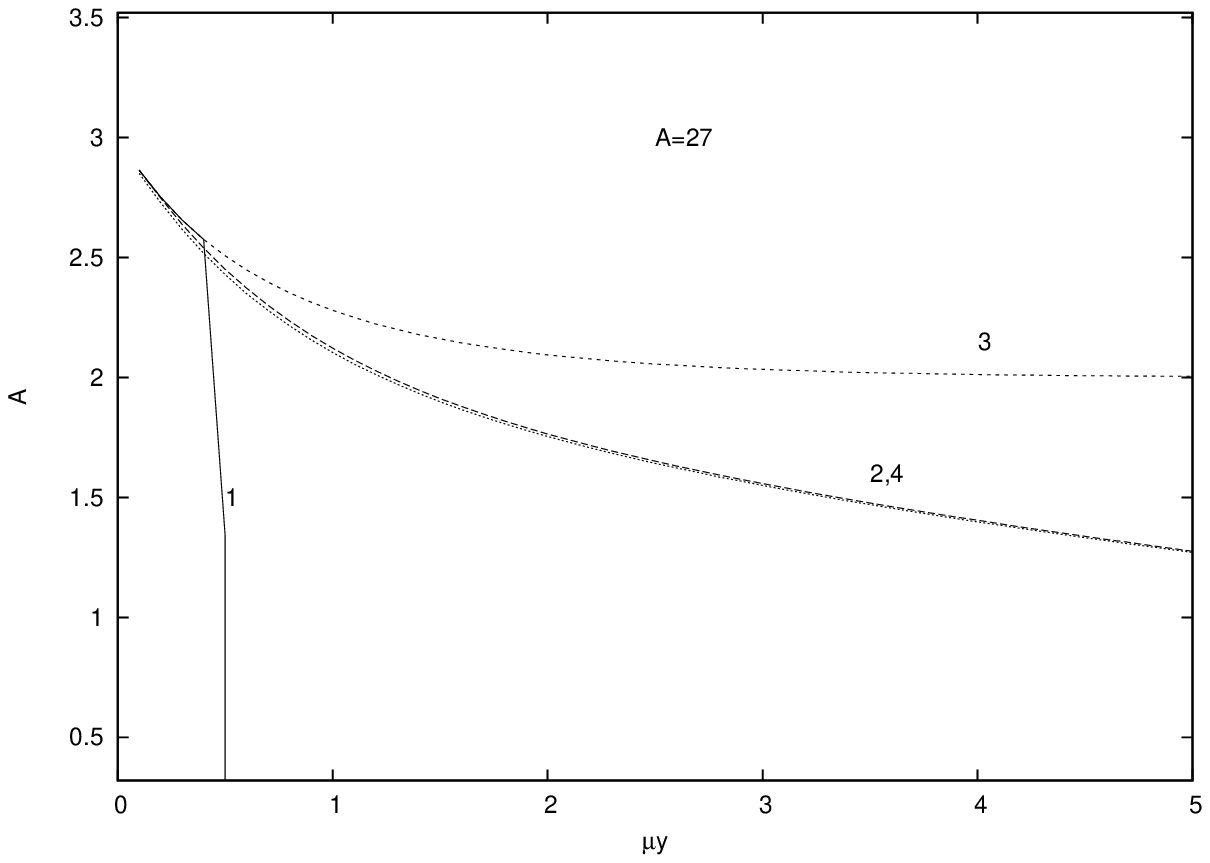}
\\
\vspace* {1 cm}
\includegraphics[width=6 cm]{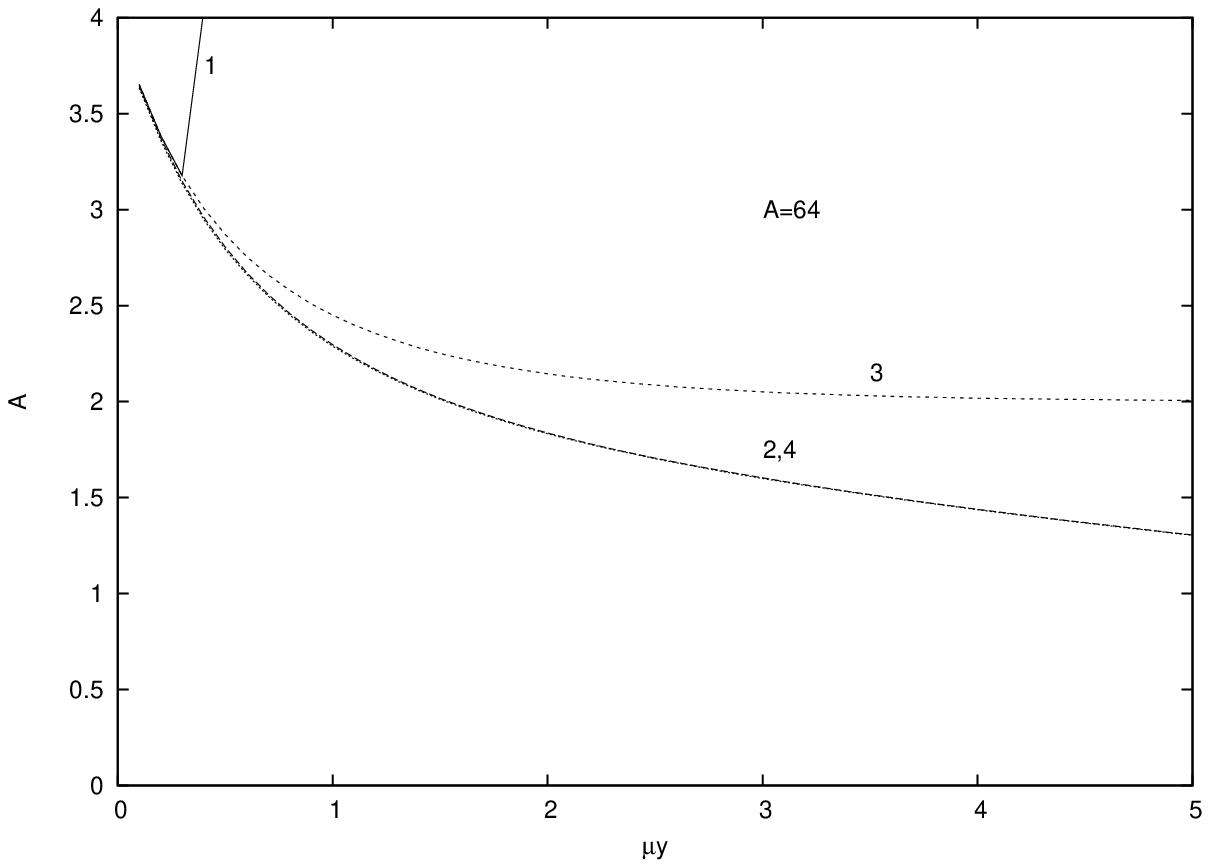}
\includegraphics[width=6 cm]{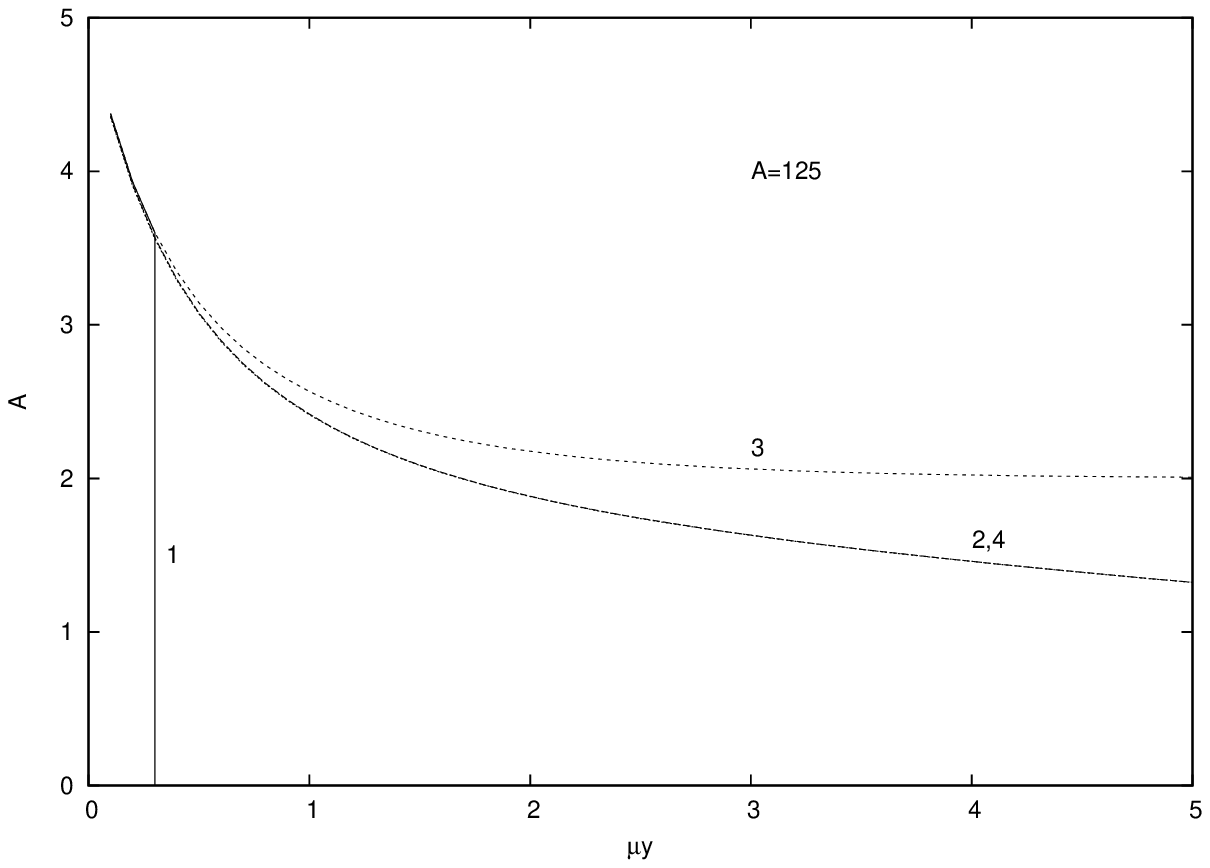}
\caption{Same as in Fig. \ref{fig1} for $\rho=2$ }
\label{fig2}
\end{figure}

\begin{figure}[ht]
\includegraphics[width=6 cm]{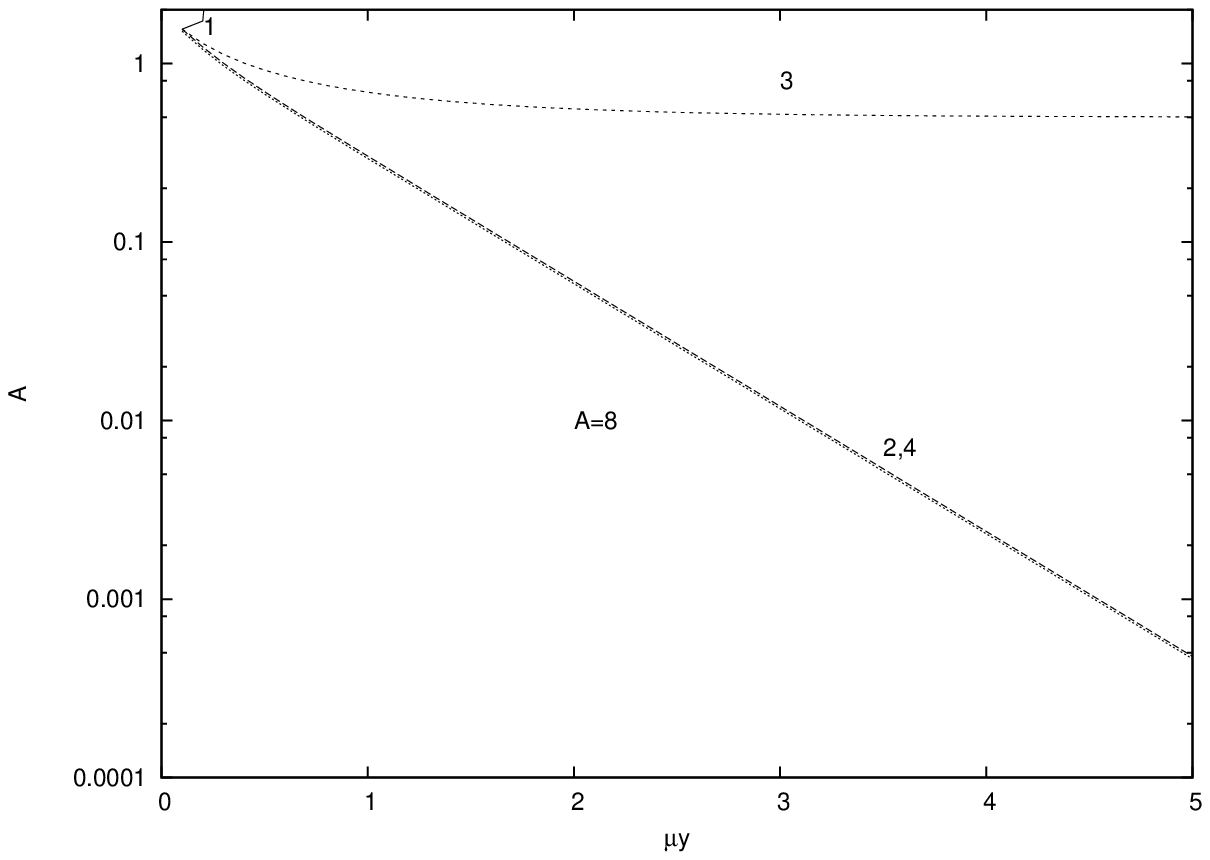}
\includegraphics[width=6 cm]{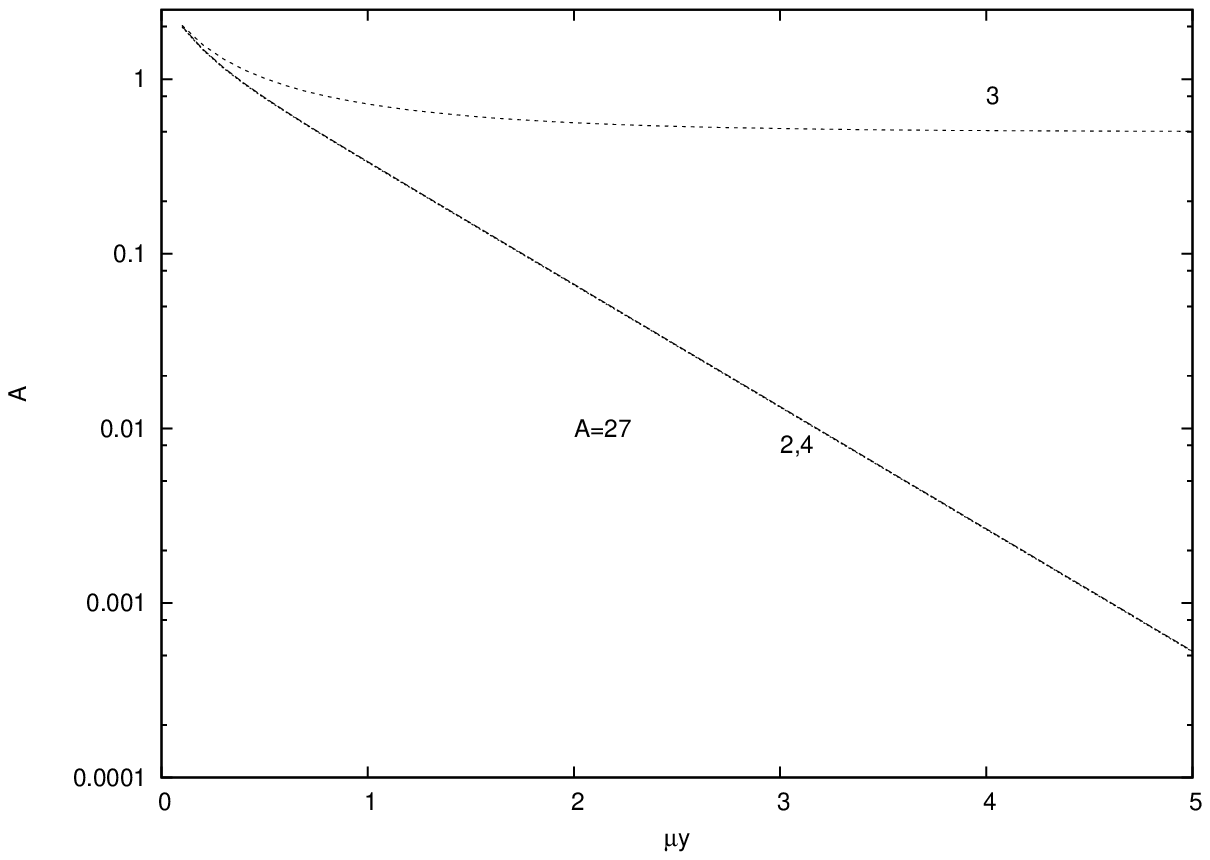}
\\
\vspace* {1 cm}
\includegraphics[width=6 cm]{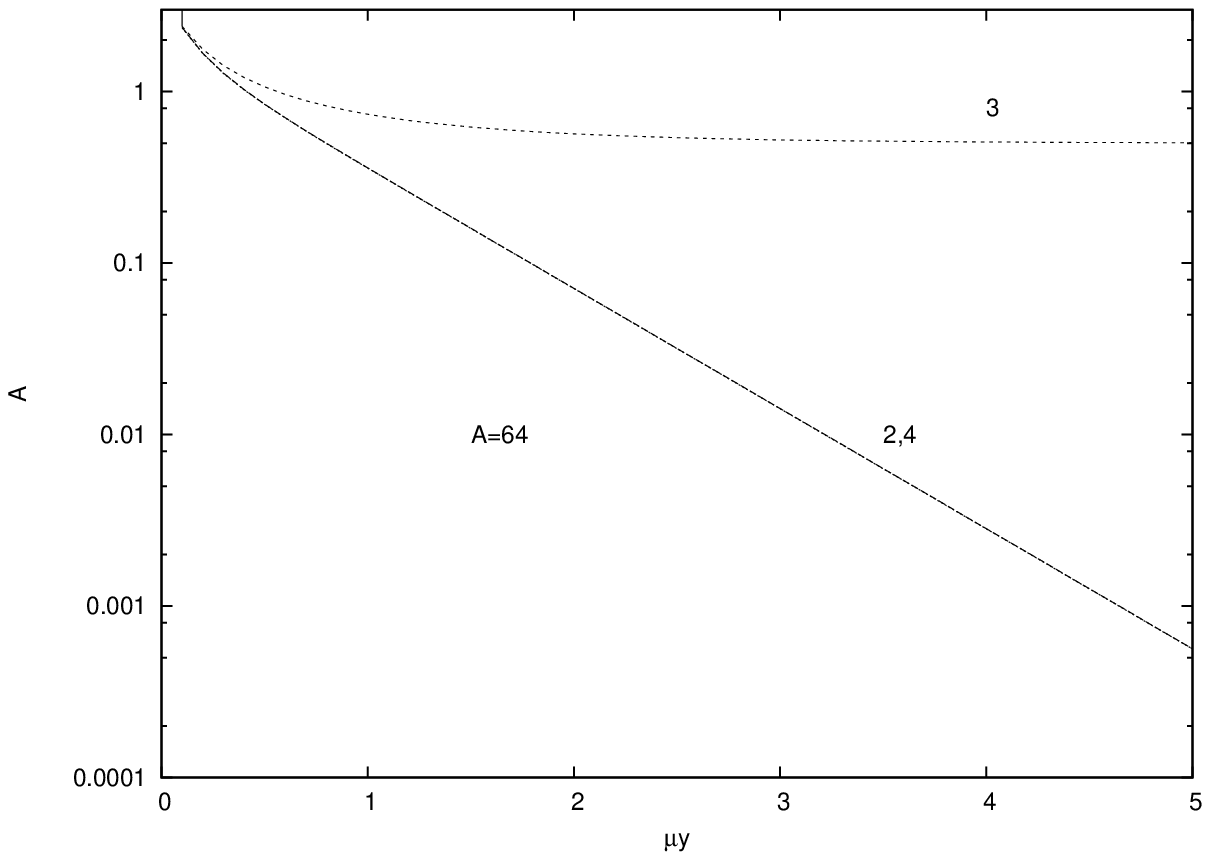}
\includegraphics[width=6 cm]{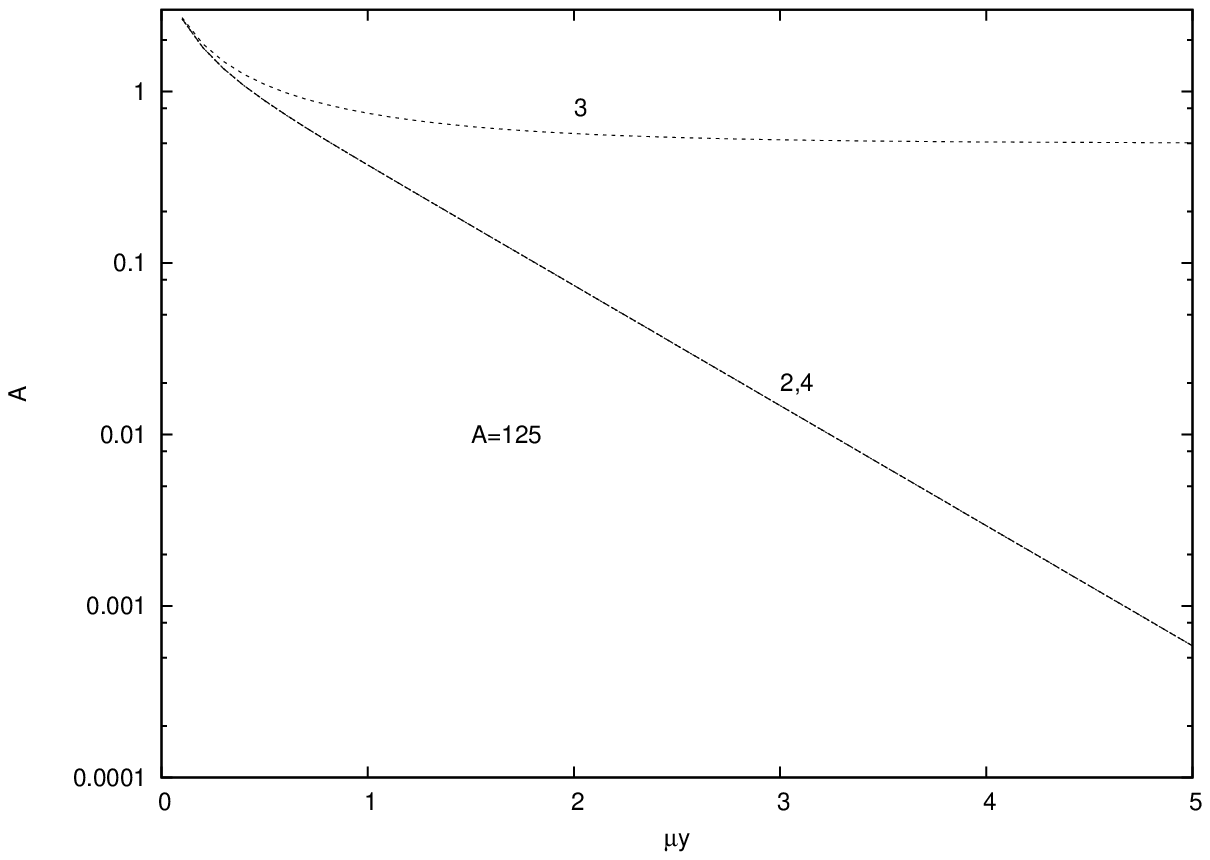}
\caption{Same as in Fig. \ref{fig1} for $\rho=0.5$}
\label{fig3}
\end{figure}

Our results for $A=8,27,64$ and 125 are shown in Figs. \ref{fig1},\ref{fig2} and \ref{fig3} for the three mentioned values
of $\rho$ respectively. In each figure we compare the fan amplitude for the finite nucleus with a given $A$ (1),
the full amplitude with loops for the finite nucleus (2), the fan amplitude for infinite nucleus Eq. (\ref{fan}) with $g_f$
given by $A^{1/3}$ (3) and finally the full amplitude with loops calculated for the infinite nucleus (with eikonal
initial function) and also with $g_f=A^{1/3}$ (4).

\section{Discussion}
Inspection of our numerical results in Figs. \ref{fig1}-\ref{fig3} leads to the following conclusions.

1) As expected, the finite sum of fan diagrams corresponding to the given nucleus with a finite atomic
numver correctly describes the hA amplitude for very small values of the triple pomeron coupling
($\rho=10$) up to a certain value of scaled rapidity $\bar{y}<\bar{y}_{max}$ after which the result sharply
blows up practically to infinity. The limiting rapidity $\bar{y}$ depends on $A$ very weakly,
diminishing from 1.5 for $A=8$ to 1.1 at $A=125$. Of course these features can be immediately read from
Eq. (\ref{schw1}). Note that at $\rho=0.5$ the interval of $\bar{y}$ where finite fans have some sense is
close to zero, so that the corresponding curve (1) is not visible in Fig. \ref{fig3}.

2) Remarkably with loop taken into account the results for finite nuclei and infinite ones practically
coincide in all cases except for $A=8$ at $\rho=10$. Moreover at comparatively high rapidities they are very weakly dependent on $A$
(but strongly dependent on $\rho$).
For instance for $\rho=2$ at $\bar{y}=5$ the hA amplitudes for $A=$8, 27, 64 and 125 are found to be
1.23, 1.28, 1.31 and 1.34 with the total difference less than 10\%. This means that the structure of
the hA amplitude with loops taken into account is due mostly to formation of loops during evolution
so that the amplitude  quickly forgets the initial state. In the exeptional case $A=8$ and $\rho=10$
the small values of $\lambda$ and $A$ evidently do not allow to form enough loops to  strongly influence the
evolving amplitude.

3) At small values of the triple pomeron coupling $\lambda$ ($\rho=10$) fan diagrams with infinite number of splitting
(the Schwimmer formula (\ref{schw})) describe the total amplitude with loops very well. This is in spite of the fact that
 with a finite nucleus infinite fans seem to have nothing to
 do with the physical amplitude. So the analytic continuation involved in extending the validity
 of the finite fans to infinite ones seems to effectively take into account contributions from loops at small enough
 $\lambda$. This surprizing result might have some bearing on the validity of the fan amplitude
 in general and in the pertubative QCD (BK eqiation) in particular.

 4) With the growth of $\lambda$  also infinite fans cease to describe the amplitude, which can be seen from
 Figs. \ref{fig2} and \ref{fig3}. With $\rho=0.5$ at $\bar{y}=5$ they overestimate the amplitude by 3 orders of magnitude.

 \section{Conclusion}
For realisitic nuclei with finite atomic numbers $A$ the standard infinite series of fan diagrams in the Regge theory
for hA amplitude
converts into a finite sum. With the elementary hadron cross-section rising at large energies this finite sum preserves
physical meaning only up to a certain maximal energy, which in fact is not very large and goes down with the rise of $A$.
Inclusion of loop diagrams may cure this situation. To study this problem we considered the toy Regge model existing
in zero transverse dimensions where the loops can be taken into account by numerical methods.

Our results first show that for finite nuclei the model with contribution from loops included gives reasonable results up to very high energies. Second we
discover that the found hA amplitude is practically identical with the one which corresponds to the infinite nucleus.
In fact at large energies the found amplitude is weakly dependent on the initial amplitude at zero energies, so that
it is formed completely from loop contributions.

Finally we found that at small enough triple pomeron coupling $\Lambda$ the infinite series of fan diagrams
gives a good description of the amplitude for finite nuclei in spite of the fact that in reality the series should be cut
and the cut series has no physical sense at large energies. The analytic continuation in parameters involved in making
the cut series to converge seems to somehow take the loop contribution into account. This conclusion, as mentioned,
could explain the success of using infinite fan diagrams in the description of hA amplitudes both in the local Regge theory
and the perturbative QCD.

 \section{Acknowledgments}
The author thanks  the Bologna Physics department and
INFN for hospitality. He also thanks  G.P.Vacca for very interesting and constructive
discussions.
The author acknowledges Saint-Petersburg State University
for a research grant 11.38.223.2015. This study has been also
supported by the RFBR grant 15-02-02097.

\end{document}